\documentclass{article}

\usepackage{arxiv}

\usepackage[utf8]{inputenc} 
\usepackage[T1]{fontenc}    
\usepackage{hyperref}       
\usepackage{url}            
\usepackage{booktabs}       
\usepackage{amsfonts}       
\usepackage{nicefrac}       
\usepackage{microtype}      
\usepackage{lipsum}		
\usepackage{graphicx}
\usepackage{natbib}
\usepackage{doi}

\title{Ética para LLMs: o compartilhamento \\ de dados sociolinguísticos}

\date{} 					

\author{ \href{https://orcid.org/0000-0002-0480-0422}{\includegraphics[scale=0.06]{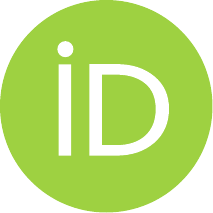}\hspace{1mm}Marta Deysiane Alves Faria Sousa} \\
	Programa de Pós-graduação em Letras\\
	Universidade Federal de Sergipe -- UFS\\
	\texttt{mpintin@gmail.com} \\
	\And
	\href{https://orcid.org/0000-0002-4972-4320}{\includegraphics[scale=0.06]{orcid.pdf}\hspace{1mm}Raquel Meister Ko. Freitag} \\
	Departamento de Letras Vernáculas\\
	Universidade Federal de Sergipe --  UFS\\
	\texttt{rkofreitag@academico.ufs.br} \\
    \And
	\href{https://orcid.org/0009-0000-5270-8033}{\includegraphics[scale=0.06]{orcid.pdf}\hspace{1mm}Túlio Sousa de Gois} \\
	Departamento de Computação\\
	Universidade Federal de Sergipe --  UFS\\
	\texttt{tuliosg@academico.ufs.br} \\
}

\date{}

\renewcommand{\headeright}{}
\renewcommand{\shorttitle}{Ética para LLMs: o compartilhamento de dados sociolinguísticos}

\begin{document}
\maketitle

\begin{abstract}
	The collection of speech data carried out in Sociolinguistics has the potential to  enhance large language models due to its quality and representativeness. In this paper, we examine the ethical considerations associated with the gathering and dissemination of such data. Additionally, we outline strategies for addressing the sensitivity of speech data, as it may facilitate the identification of informants who contributed with their speech.
\end{abstract}


\section{Introdução}

O uso da inteligência artificial (IA) tem se tornado cada vez mais presente na vida da população brasileira. Particularmente, o uso de IA na educação é uma realidade \citep{de2023desafios, leao2021inteligencia} a qual, como educadores, tivemos que nos adaptar rapidamente tentando extrair o melhor delas para tornar o processo de ensino e aprendizagem mais interessante. Contudo, o uso dessas tecnologias em quaisquer áreas do conhecimento perpassa por questões éticas cujo debate no cenário brasileiro ainda é incipiente.

Em 2023, a Confederação de Organizações Europeias de Proteção de Dados reconheceu que a Inteligência Artificial Gerativa (IAG) é uma tecnologia muito nova e que as organizações que cuidam da proteção de dados devem remodelar suas normativas para atender às diferentes demandas que estão surgindo junto com a incorporação desse tipo de tecnologia em diversos espaços. Isso se deve ao fato de que os modelos de IAG são treinados em grandes volumes de dados, em sua maioria mal documentados, codificando, como consequência, viéses, estereótipos, discursos nocivos, além de os reproduzir em suas respostas \citep{bender2021dangers}. 

No Brasil, o Governo Federal elaborou a Proposta de Plano Brasileiro de Inteligência Artificial 2024-2028, que prevê o investimento de 23 bilhões de reais. Sob a premissa de uma ``IA para o bem de todos", são destacadas visões como a centralidade no ser humano, prevenindo desigualdade e viéses, e a transparência e responsabilidade, garantindo a privacidade e a sobreania dos dados. Contudo, apenas 0,45\% do orçamento é destinado ao ``Apoio ao Processo Regulatório e de Governança da IA", evidenciando ainda mais a necessidade de discussões urgentes sobre a ética na IA no Brasil\footnote{Proposta de Plano Brasileiro de Inteligência Artificial 2024-2028. Disponível em: https://www.gov.br/mcti/pt-br/acompanhe-o-mcti/noticias/2024/07/plano-brasileiro-de-ia-tera-supercomputador-e-investimento-de-r-23-bilhoes-em-quatro-anos/ia\_para\_o\_bem\_de\_todos.pdf/view}.

A preocupação ética na alimentação de grandes modelos de linguagem (large language models - LLMs) atravessa também o trabalho de linguistas. Como já dito, essas tecnologias precisam de grandes volumes de dados, no caso de chatbots, por exemplo, dados linguísticos são necessários para que as máquinas possam ter um fluxo de fala (ou escrita) similar ao de seres humanos, colocando aos linguistas o desafio de pensar em como proceder para proteger eticamente os dados colhidos. Neste artigo, então, discutimos desde a base de coleta de dados linguísticos ao compartilhamento desses dados para alimentar LLMs.

\section{Regulamentação} 

Nos últimos anos, muito tem-se debatido sobre Ciência Aberta e a centralidade dos dados de pesquisas no fortalecimento do conhecimento produzido pela academia, de tal forma que os periódicos científicos têm demandado dos pesquisadores acesso aos dados que geraram as publicações. Isso se deve ao fato de que a maior transparência na condução, compartilhamento de dados e publicação das pesquisas acarreta não só maior fortalecimento do rigor metodológico e experimental, mas também a confiabilidade na produção científica por parte do público geral \citep{lyon2016transparency}. Ademais, no escopo do cenário científico brasileiro, no qual grande parte do financiamento de projetos de pesquisa parte da iniciativa pública, é esperado que os dados coletados sejam de domínio público. Entretanto, existem aspectos concernetes à responsabilização das instituições e dos pesquisadores com os dados fornecidos pelos indivíduos que aceitam participar das pesquisas que devem balizar o compartilhamento desses dados \citep{freitag2021linguistic, meister2022sociolinguistic}.

A normativa que trata das pesquisas com seres humanos, no Brasil, é a Resolução Nº 510, de 07 de abril de 2016 do Conselho Nacional de Saúde. Destacamos o fato de que seu artigo 9º apresenta os direitos dos participantes das pesquisas, evidenciando o protagonismo destes na tomada de decisão quanto às informações que podem ser publicadas e tornadas públicas. Ainda, neste documento, encontramos informações acerca da documentação que deve ser entregue aos participantes para torná-los cientes dos possíveis ganhos e danos que possam ser causados em decorrência da pesquisa em curso.

Além da Resolução Nº 510, de 07 de abril de 2016 do Conselho Nacional de Saúde, no Brasil, existe a Lei Geral de Proteção de Dados (Lei 13.709, de 2018, doravante LGPD), inspirada na Regulamento Geral de Proteção de Dados da Europa. Essa lei tem como objetivo assegurar aos cidadãos brasileiros o controle de suas informações pessoais, demandando daqueles que as coletam o pedido de consentimento sobre os usos desses dados bem como a escolha dos usuários sobre as formas de visualização, correção e também exclusão de dados pessoais que circulam na internet.

Ao pensarmos, então, em coleta de dados, principalmente, no caso de pesquisadores da área de Linguística que produzem \textit{corpora} para suas pesquisas, devemos observar essas duas regulamentações para conduzirmos nossos estudos. Assim, os procedimentos éticos de entrada em campo e de compartilhamento de dados devem refletir essas normativas e assegurar aos participantes o controle das informações que eles oferecem.

\section{Dados sociolinguísticos}

No escopo da Sociolinguística, a compilação de bancos de dados se dá principalmente por meio de entrevistas sociolinguísticas visto que elas se apresentam como o melhor instrumento para se capturar a língua vernacular em grandes volumes (entre uma e duas horas de duração) e com melhor qualidade em termos de gravação \citep{labov1981field}. Assim, os bancos de dados sociolinguísticos, que já são utilizados no treinamento de ferramentas de PLN \citep{sousa2024bancos}, se tornam produtos com alto valor para a área de IA por possuírem qualidade em termos de gravação, transcrição e quantidade para alimentar LLMs.

É inevitável, então, que os dados de fala coletados preservem as condições da fala natural do indivíduo, tornando-se um paradoxo entre a preservação da identidade dos participantes e o reconhecimento de suas vozes. Para mitigiar os efeitos desse paradoxo, o termo de consentimento e as licenças de uso são fundamentais \citep{calamai2018fair, mello2021trabalhando}.

\subsection{Caminhos a seguir}

Primeiramente, o termo de consentimento deve prever que o consentimento para utilização dos dados pode ser interrompido em qualquer estágio da pesquisa e indicar o local em que os dados serão depositados. Ademais, os participantes devem ser informados sobre o possível compartilhamento desses dados, determinando seu tipo circulação e a finalidade desse compartilhamento.

Ademais, visando garantir ainda mais o controle dos dados compartilhados tanto por parte dos pesquisadores quanto dos participantes, é importante a escolha adequada de licenças de uso. Existem dois tipos de licenças para garantir os direitos do autor, a licença GNU General Public License (GNU GPL) e as Creative Commons (CC). A licença GNU GPL é menos restritiva que as CC em termos de reutilização, sendo a única restrição a de que os derivados produzidos com os dados sejam de acesso aberto. Está no bojo das licenças CC respeitar os direitos autorais e conexos.

É no contexto do compartilhamento de dados de fala que uma licença adequada contribui tanto para a atribuição da devida autoria daquele produto intelectual quanto em uma maior transparência das pesquisas na área. Nesse caso, sugerimos ``Atribuição-NãoComercial-CompartilhaIgual", uma licença CC que, por meio dela, se protege a autoria e é autorizado o desenvolvimento de novas tecnologias a partir dos dados disponibilizados. Assim, o licenciante (o criador do banco de dados) permite que o  licenciado (quem usará o conteúdo) utilize e faça novos trabalhos com os dados gerados, contanto que cite os licenciantes e utilize a mesma licença na criação dos derivados para propósitos não-comerciais.

\section{Considerações finais}

Neste artigo, abordamos de maneira abrangente a necessidade de se considerar questões éticas associadas à coleta e uso de dados de fala na Sociolinguística para a alimentação de LLMs. Discutimos o dilema de se preservar a identidade dos informantes bem como a utilidade desses dados para IA. Sugerimos o uso de termos de consentimento e licenças de uso como passos importantes para garantir que os dados sejam coletados e compartilhados de maneira ética e responsável.
A continuidade das discussões sobre ética na IA no Brasil é urgente, principalmente devido ao seu rápido avanço e assimilação pela sociedade. A implementação efetiva das regulamentações existentes e a promoção de uma cultura de responsabilidade ética serão essenciais para o desenvolvimento sustentável e responsável da IA em nosso país.

\bibliographystyle{unsrtnat}
\bibliography{references}  

\end{document}